\documentclass[preprint,aps,draft]{revtex4}

\usepackage{graphicx}
\usepackage{dcolumn}
\usepackage{bm}

\begin{document}

\setlength{\unitlength}{1mm}
\textwidth 15.0 true cm
\textheight 22.0 true cm
\headheight 0 cm
\headsep 0 cm
\topmargin 0.4 true in
\oddsidemargin 0.25 true in

\title{Dark Energy with $w>-4/3$.}

\author{Andrei Gruzinov}
 \affiliation{Center for Cosmology and Particle Physics, Department of Physics, New York University, NY 10003}

\date{May 5, 2004}

\begin{abstract}

Acceleration of the universe might be driven by a continuous elastic medium -- elastic dark energy (Bucher and Spergel 1999). Elastic dark energy can stably support equations of state with pressure to energy ratio $w\equiv p/\epsilon > -4/3$. Stable expansion with $w<-1$ combined with an assumption of ``reheating of elastic energy'' leads to exotic possibilities such as Expanding Cyclic Universe -- an ever-expanding universe with periodically repeating  inflationary epochs.

\end{abstract}


\maketitle

\section{Introduction}

According to observations (see \cite{teg} and references therein) dark energy has pressure to energy ratio $w\equiv p/\epsilon \approx -1$. During inflation one must also have $w\approx -1$. It is therefore natural to assume that the current epoch is the beginning of the new inflationary period. If this new inflation is to repeat the previous one, the dark energy density must grow, which requires $w<-1$. Equations of state with $w<-1$ are allowed by an elastic medium.

Relativistic elastic continuous medium can be responsible for both the current acceleration of the universe \cite{buch} and for inflation \cite{gruz}. Elastic dark energy can be stable for $-4/3<w<1$. Assume that $w$ is slightly smaller than $-1$ during the current epoch. Then the dark energy is slowly growing, while matter and radiation are rapidly diluted. The sub-horizon universe will become homogeneous and isotropic almost everywhere. When elastic energy becomes too large, inflation might be terminated by reheating, when most of elastic energy is converted into radiation. A small amount of elastic energy remains, and after a while elastic energy will come to dominate, and a new inflationary period will commence. This is similar to the cyclic picture (see \cite{stein} and references therein), although in our model the universe keeps expanding at all time. 

Unfortunately, we have nothing to say about the naturalness of elastic inflation, that is why  $w\approx -1$ if any $-4/3<w<1$ is theoretically allowed. The assumption that elastic reheating converts nearly all, but not all elastic energy into radiation is also highly unnatural. Still, the theoretical possibility of a stable medium with $w<-1$, and the related possibility of an Expanding Cyclic Universe seem worth noting.

We describe the equation of state and stability of relativistic elastic medium in \S2. A model of an Expanding Cyclic Universe is discussed in \S3. We summarize in \S4. 

\section{Elastic Medium}
Continuous elastic medium is a collection of time-like world lines filling space-time. Elastic energy density is an arbitrary function of spatial distances between neighboring world lines (see \cite{buch} and references therein). The cosmological perturbation theory of an elastic medium has been developed in \cite{buch, gruz}. For our current purposes, it is sufficient to consider subhorizon perturbations. If the subhorizon perturbations have propagation speeds $\sim 1$, gravity will not affect the subhorizon stability. The super-horizon evolution is included in the calculation of elastic inflation  \cite{gruz} and will not concern us here. The elastic energy-momentum tensor is written down in \cite{gruz}, here we use the results of that paper without further references.

The subhorizon perturbations are longitudinal (scalar) and transverse (vector) sound waves. Let $c_s$ be the speed of scalar sound of the elastic medium, and $c_v$ -- the speed of vector sound, $\epsilon _\Lambda$ -- the elastic energy density, $p_\Lambda$ -- the elastic pressure. The unperturbed equation of state of elastic medium, relating the change of energy and pressure in a uniformly expanding universe, reads
\begin{equation}
{dp_\Lambda \over d\epsilon_\Lambda } =c_s^2-{4\over 3}c_v^2.
\end{equation}
The subhorizon stability and causality require 
\begin{equation}
0<c_s^2<1, ~~~0<c_v^2<1.
\end{equation}
These requirements constrain elastic equations of state. Assuming, for simplicity, constant sound speeds, we get a possible range of pressure to energy ratios
\begin{equation}
-{4\over 3}<w<1.
\end{equation}

We are mostly interested in the case $w\approx -1$. Then perturbations of the energy-momentum tensor of elastic medium are very small (perturbations of $T^\mu _\nu=\Lambda \delta ^\mu _\nu$ are equal to zero). The gravitational interaction of elastic medium with normal matter are then negligible. We also assume that there are no other interactions.

\section{Expanding Cyclic Universe}

Consider a model universe consisting  only of elastic medium (energy $\epsilon _\Lambda$, pressure $p_\Lambda=w\epsilon _\Lambda$, with $w$ close to but less than $-1$) and radiation (energy $\epsilon _r$, pressure $p_r=\epsilon _r/3$). A realistic model, with late-time matter domination can be built in a similar way. The heart of the problem -- the reheating of elastic energy -- is not affected by this simplifying assumption.

The evolution of elastic universe with radiation is given by the following Friedmann equations (flat FRW):
\begin{equation}\label{s1}
\dot{\epsilon _\Lambda }=(-3w-3)H\epsilon _\Lambda -F(\epsilon _\Lambda ,\epsilon _r),
\end{equation}
\begin{equation}
\dot{\epsilon _r }=-4H\epsilon _r +F(\epsilon _\Lambda ,\epsilon _r),
\end{equation}
\begin{equation}\label{s2}
M_{Pl}^2H^2=\epsilon _\Lambda +\epsilon _r.
\end{equation}
Here $H$ is the Hubble constant, $M_{Pl}$ is the Planck mass, $F$ is the reheating rate. 

The system (\ref{s1}-\ref{s2}) has equilibrium points
\begin{equation}
(-3w-3)H\epsilon _\Lambda =4H\epsilon _r=F(\epsilon _\Lambda ,\epsilon _r).
\end{equation}
For some reheating rates, the equilibrium points might be unstable. Then (\ref{s1}-\ref{s2}) can have a limit cycle solution. Starting from an arbitrary initial condition, the system approaches the limit cycle. On the limit cycle, the solution is periodic as described in the introduction: (i) inflation is terminated by reheating, (ii) reheating leaves behind a trace amount of elastic energy, (iii) elastic energy finally comes to dominate, initiating a new inflationary cycle.

\section{Discussion}

Elastic Dark Energy can have an equation of state with $w<-1$. This may allow a growth of elastic energy, which may lead to a new inflation and big bang.

\begin{acknowledgments}
This work was supported by the David and Lucile Packard Foundation.
\end{acknowledgments}

\begin{appendix}

\end{appendix}

\end{document}